\documentclass[11pt]{article}
\usepackage{amssymb}
\usepackage{amsmath}
\usepackage{amsfonts}
\usepackage{macro}
\usepackage{graphicx}

\title{\Large\bf Two-way Quantum One-counter Automata}
\author{
  {\large $\mbox{\bf Tomohiro Yamasaki}^{\ast}$}
  \and
  {\large $\mbox{\bf Hirotada Kobayashi}^{\ast \ddagger}$}\\
  {\tt \hspace*{-30ex} \{yamasaki, hirotada, imai\}@is.s.u-tokyo.ac.jp \hspace*{-20ex}}
  \and
  {\large $\mbox{\bf Hiroshi Imai}^{\dagger \ddagger}$}
}

\date{2 October 2001}
\begin{document}

\maketitle

%\vspace*{-7ex}
%\begin{center}
%\end{center}

\footnotetext[1]{
Department of Information Science,
        Graduate School of Science, The University of Tokyo,
        7-3-1 Hongo, Bunkyo-ku, Tokyo 113-0033, Japan.}
\footnotetext[2]{
Department of Computer Science,
        Graduate School of Information Science and Technology,
        The University of Tokyo,
        7-3-1 Hongo, Bunkyo-ku, Tokyo 113-0033, Japan.}
\footnotetext[3]{
Quantum Computation and Information Project, ERATO, JST,
        5-28-3 Hongo, Bunkyo-ku, Tokyo 113-0033, Japan.}

\begin{abstract}

After the first treatments of quantum finite state automata
by Moore and Crutchfield and by Kondacs and Watrous,
a number of papers study the power of quantum finite state automata
and their variants.
This paper introduces
a model of two-way quantum one-counter automata (2Q1CAs),
combining the model of two-way quantum finite state automata (2QFAs)
by Kondacs and Watrous
and the model of one-way quantum one-counter automata (1Q1CAs)
by Kravtsev.
We give the definition of 2Q1CAs with well-formedness conditions.
It is proved that
2Q1CAs are at least as powerful as
classical two-way deterministic one-counter automata (2D1CAs),
that is, 
every language $L$ recognizable by 2D1CAs
is recognized by 2Q1CAs with no error.
It is also shown that several non-context-free languages
including $\{a^n b^{n^2} \mid n \ge 1\}$
and $\{a^n b^{2^n} \mid n \ge 1\}$ are recognizable
by 2Q1CAs with bounded error.
 
\end{abstract}

\section{Introduction}
Quantum finite state automata were introduced by
Moore and Crutchfield \cite{MooreC1997}
and Kondacs and Watrous \cite{KondacsW1997} independently.
In particular, the latter introduced
both of the quantum counterparts
of one-way and two-way finite state automata,
which are called one-way and two-way quantum finite state automata
(1QFAs and 2QFAs), respectively.
In the classical case, it is well known that
two-way deterministic finite state automata (2DFAs)
remain the same in view of language recognition
as one-way deterministic finite state automata (1DFAs),
and both of the class of languages recognized by 1DFAs and 2DFAs
are exactly the class of regular languages.
In the quantum case, the situation becomes different.
In fact, Kondacs and
Watrous \cite{KondacsW1997} showed that the class
of languages recognized by 1QFAs is properly contained in the class of regular
languages, while that of 2QFAs properly contains the class of regular
languages.
Subsequently,
a number of papers
study the power of quantum finite state automata
\cite{AmbainisF1998, AmanoI1999, AmbainisNTV1999, BrodskyP1999,
  AmbainisBFK1999, Nayak1999, AblayevG2000, AmbainisKV2001, AmbainisK2001}
and their variants
\cite{AmbainisBFGK1999, AmbainisW1999, Kravtsev1999, YamasakiKTI2000,
  Golovkins2000, BonnerFK2001, FreivaldsW2001}.

As for the counter automata,
Kravtsev \cite{Kravtsev1999} introduced a quantum version of
one-way one-counter automata (one-way quantum one-counter automata,
1Q1CAs). In the classical case, it can be easily seen that
the class of languages
recognized by one-way deterministic one-counter automata (1D1CAs)
is properly contained in the class of context-free languages (CFL),
because 1D1CAs are the special case of one-way pushdown automata
(1PDAs) and the class of languages recognized by 1PDAs is equal to that 
of CFL.
Kravtsev showed that some non-regular languages and some
non-context-free languages are recognized by 1Q1CAs with bounded error,
which means,
in some cases, 1Q1CAs are more powerful than the classical counterparts.
A consecutive work by Yamasaki, Kobayashi, Tokunaga, and Imai \cite{YamasakiKTI2000} showed that
these non-regular or non-context-free languages are also recognized by
one-way probabilistic reversible one-counter automata (1PR1CAs) with
bounded error.
They also showed that 
there exists a regular language which cannot be recognized by
bounded error 1Q1CAs, which means that, in some cases, 1Q1CAs are less powerful
than the classical counterparts.
Recently Bonner, Freivalds, and Kravtsev \cite{BonnerFK2001} showed that
there exists a language which cannot be recognized by
one-way probabilistic one-counter automata (1P1CAs) with bounded error,
but can be recongnized by bounded error 1Q1CAs.

In one-way case, the automaton can read each symbol of an input only
once. This can be regarded as a quite strong restriction
for quantum cases, because each
step of evolution of quantum automata should be unitary.
If this restriction of one-way is removed, how much power is added to
quantum one-counter automata?
From this point of view, this paper introduces
two-way quantum one-counter automata (2Q1CAs)
and analyzes the power of this model.
We give a formal definition of 2Q1CAs with well-formedness conditions,
and compare the power of 2Q1CAs
with that of two-way deterministic one-counter automata (2D1CAs).
It is shown that if a language $L$ is recognized by 2D1CAs,
$L$ is necessarily recognized by 2Q1CAs with no error,
which means that 2Q1CAs are at least as powerful as 2D1CAs.
It is also shown that
bounded error 2Q1CAs can recognize some non-context-free languages such as
$\{a^n b^{n^2} \mid n \ge 1\}$, $\{a^m b^n c^{mn} \mid m, n \ge 1\}$,
$\{a^n b^{2^n} \mid n \ge 1\}$, and so on.
These languages do not seem to be recognized by 2D1CAs, therefore,
the authors conjecture that 2Q1CAs are strictly more powerful than 2D1CAs.

The remainder of this paper is organized as follows.
In Section 2 we give a formal definition of 2Q1CAs with
well-formedness conditions and introduce the concept of simple 2Q1CAs.
In Section 3, using the technique of reversible simulation,
we prove that if a language $L$ is recognized by 2D1CAs,
$L$ is necessarily recognized by 2Q1CAs with no error.
In Section 4, we show that 2Q1CAs can recognize
some non-context-free languages with bounded error.
Finally we conclude with Section 5,
which summarizes our results and mentions several open problems.

\section{Definitions}
\label{def}

\subsection{Two-way One-counter Automata}

Here we give formal definitions of two-way one-counter
automata both in classical and quantum cases.
First we define two-way deterministic one-counter automata.

\begin{df}
 A two-way deterministic one-counter automaton (2D1CA) is defined as
 $M = (Q, \Sigma, \delta, q_0, \Qacc, \Qrej)$, where
 $Q$ is a finite set of states,
 $\Sigma$ is a finite input alphabet,
 $q_0$ is an initial state,
 $\Qacc \subset Q$ is the set of accepting states,
 $\Qrej \subset Q$ is the set of rejecting states, and
 $\delta : Q \times \Gamma \times S \rightarrow
 Q \times \{-1, 0, +1\} \times \{\leftarrow, \downarrow, \rightarrow\}$
 is a transition function, where
 $\Gamma = \Sigma \cup \{\cent, \dollar\}$,
 symbol $\cent   \not\in \Sigma$ is the left  end-marker,
 symbol $\dollar \not\in \Sigma$ is the right end-marker, and
 $S = \{0, 1\}$.
\end{df}

We assume that each 2D1CA has a counter which can contain an arbitrary
integer. When $\delta$ returns $-1, 0, +1$, respectively, as the second
element, the automaton decreases the counter value by 1, retains
the same, and increases by 1. Let $s = \sign(k)$, where $k$ is the
counter value and $\sign(k) = 0$ if $k = 0$, otherwise $1$. 

We also assume that every input written on the tape is started by
$\cent$ and terminated by $\dollar$, and that such a tape is circular.
When $\delta$ returns $\leftarrow, \downarrow,
\rightarrow$, respectively, as the third element,
the automaton moves its tape head left by a
square, remains stationary, and moves right by a square.

At the beginning of computation, the state is $q_0$, the counter value
is $0$, and the tape head is scanning the left-most symbol of an input.
At each step, the automaton reads a symbol $\sigma$ in the state $q$,
checks whether the counter value is $0$ or not (i.e. checks $s$), and
finds an appropriate transition $\delta(q, \sigma, s) = (q^\prime, c,
d)$, where $c \in \{-1, 0, +1\}$ and $d \in \{\leftarrow, \downarrow, 
\rightarrow\}$. Then it enters the state $q^\prime$,
updates the counter value according to $c$,
and moves its tape head in direction $d$.
The automaton accepts the input if it enters the final state in $\Qacc$ and
rejects if it enters the final state in $\Qrej$.

Next we define two-way quantum one-counter automata.

\begin{df}
 A two-way quantum one-counter automaton (2Q1CA) is defined as
 $M = (Q, \Sigma, \delta, q_0, \Qacc, \Qrej)$, where
 $Q, \Sigma, q_0, \Qacc, \Qrej$ are defined in the same manner
 as those with 2D1CAs.
 A transition function $\delta$ is defined as
 $Q \times \Gamma \times S \times
 Q \times \{-1, 0, +1\} \times \{\leftarrow, \downarrow, \rightarrow\}$,
 where $\Gamma, \cent, \dollar, S$ are also defined in the same manner
 as those with 2D1CAs.
\end{df}

The definition of a counter and a tape also remains the same as those with 2D1CAs.

For any $q_1, q_2 \in Q$, $\sigma, \sigma_1, \sigma_2 \in \Gamma$,
$c \in \{-1, 0, +1\}$, $d \in \{\leftarrow, \downarrow, \rightarrow\}$,
$\delta$ satisfies the following conditions ({\it well-formedness
conditions}):
{\small
\begin{eqnarray}
     &\hspace*{-10mm}\displaystyle\sum_{q^\prime, c, d}
      \delta^*(q_1, \sigma  , s  , q^\prime,  c, d)
      \delta  (q_2, \sigma  , s  , q^\prime,  c, d)
       = \begin{cases}
          1 & q_1 =   q_2, \\
          0 & q_1 \ne q_2,
         \end{cases}~~~& \\
     &\hspace*{-10mm}\displaystyle\sum_{q^\prime, c}\Bigl(
      \delta^*(q_1, \sigma_1, s  , q^\prime,  c, \leftarrow)
      \delta  (q_2, \sigma_2, s  , q^\prime,  c, \downarrow)
  +
      \delta^*(q_1, \sigma_1, s  , q^\prime,  c, \downarrow)
      \delta  (q_2, \sigma_2, s  , q^\prime,  c, \rightarrow)
       \Bigr) = 0,~~~& \\
     &\hspace*{-10mm}\displaystyle\sum_{q^\prime, c}
      \delta^*(q_1, \sigma_1, s  , q^\prime,  c, \leftarrow)
      \delta  (q_2, \sigma_2, s  , q^\prime,  c, \rightarrow)
       = 0,~~~& \\
     &\hspace*{-10mm}\displaystyle\sum_{q^\prime, d}\Bigl(
      \delta^*(q_1, \sigma  , s_1, q^\prime, -1, d)
      \delta  (q_2, \sigma  , s_2, q^\prime,  0, d)
  +
      \delta^*(q_1, \sigma  , s_1, q^\prime,  0, d)
      \delta  (q_2, \sigma  , s_2, q^\prime, +1, d)
       \Bigr) = 0,~~~& \\
     &\displaystyle\sum_{q^\prime}\Bigl(
      \delta^*(q_1, \sigma_1, s_1, q^\prime, -1, \leftarrow)
      \delta  (q_2, \sigma_2, s_2, q^\prime,  0, \downarrow)
  +
      \delta^*(q_1, \sigma_1, s_1, q^\prime, -1, \downarrow)
      \delta  (q_2, \sigma_2, s_2, q^\prime,  0, \rightarrow)
      ~~~~~~~~~~~~~~~~~~~~~&\nonumber \\
 &{} +
      \delta^*(q_1, \sigma_1, s_1, q^\prime,  0, \leftarrow)
      \delta  (q_2, \sigma_2, s_2, q^\prime, +1, \downarrow)
  +
      \delta^*(q_1, \sigma_1, s_1, q^\prime,  0, \downarrow)
      \delta  (q_2, \sigma_2, s_2, q^\prime, +1, \rightarrow)
       \Bigr) = 0,~~~& \\
     &\displaystyle\sum_{q^\prime}\Bigl(
      \delta^*(q_1, \sigma_1, s_1, q^\prime, -1, \rightarrow)
      \delta  (q_2, \sigma_2, s_2, q^\prime,  0, \downarrow)
  +
      \delta^*(q_1, \sigma_1, s_1, q^\prime, -1, \downarrow)
      \delta  (q_2, \sigma_2, s_2, q^\prime,  0, \leftarrow)
      ~~~~~~~~~~~~~~~~~~~~~&\nonumber \\
 &{} +
      \delta^*(q_1, \sigma_1, s_1, q^\prime,  0, \rightarrow)
      \delta  (q_2, \sigma_2, s_2, q^\prime, +1, \downarrow)
  +
      \delta^*(q_1, \sigma_1, s_1, q^\prime,  0, \downarrow)
      \delta  (q_2, \sigma_2, s_2, q^\prime, +1, \leftarrow)
       \Bigr) = 0,~~~& \label{6}\\
     &\hspace*{-10mm}\displaystyle\sum_{q^\prime}\Bigl(
      \delta^*(q_1, \sigma_1, s_1, q^\prime, -1, \leftarrow)
      \delta  (q_2, \sigma_2, s_2, q^\prime,  0, \rightarrow)
  +
      \delta^*(q_1, \sigma_1, s_1, q^\prime,  0, \leftarrow)
      \delta  (q_2, \sigma_2, s_2, q^\prime, +1, \rightarrow)
       \Bigr) = 0,~~~& \label{7}\\
     &\hspace*{-10mm}\displaystyle\sum_{q^\prime}\Bigl(
      \delta^*(q_1, \sigma_1, s_1, q^\prime, -1, \rightarrow)
      \delta  (q_2, \sigma_2, s_2, q^\prime,  0, \leftarrow)
  +
      \delta^*(q_1, \sigma_1, s_1, q^\prime,  0, \rightarrow)
      \delta  (q_2, \sigma_2, s_2, q^\prime, +1, \leftarrow)
       \Bigr) = 0,~~~&\\
     &\hspace*{-10mm}\displaystyle\sum_{q^\prime, d}
      \delta^*(q_1, \sigma  , s_1, q^\prime, -1, d)
      \delta  (q_2, \sigma  , s_2, q^\prime, +1, d)
       = 0,~~~& \label{9}\\
     &\hspace*{-10mm}\displaystyle\sum_{q^\prime}\Bigl(
      \delta^*(q_1, \sigma_1, s_1, q^\prime, -1, \leftarrow)
      \delta  (q_2, \sigma_2, s_2, q^\prime, +1, \downarrow)
  +
      \delta^*(q_1, \sigma_1, s_1, q^\prime, -1, \downarrow)
      \delta  (q_2, \sigma_2, s_2, q^\prime, +1, \rightarrow)
       \Bigr) = 0,~~~& \\
     &\hspace*{-10mm}\displaystyle\sum_{q^\prime}\Bigl(
      \delta^*(q_1, \sigma_1, s_1, q^\prime, -1, \rightarrow)
      \delta  (q_2, \sigma_2, s_2, q^\prime, +1, \downarrow)
  +
      \delta^*(q_1, \sigma_1, s_1, q^\prime, -1, \downarrow)
      \delta  (q_2, \sigma_2, s_2, q^\prime, +1, \leftarrow)
       \Bigr) = 0,~~~& \\
     &\hspace*{-10mm}\displaystyle\sum_{q^\prime}
      \delta^*(q_1, \sigma_1, s_1, q^\prime, -1, \leftarrow)
      \delta  (q_2, \sigma_2, s_2, q^\prime, +1, \rightarrow)
       = 0,~~~& \\
     &\hspace*{-10mm}\displaystyle\sum_{q^\prime}
      \delta^*(q_1, \sigma_1, s_1, q^\prime, -1, \rightarrow)
      \delta  (q_2, \sigma_2, s_2, q^\prime, +1, \leftarrow)
       = 0.~~~&
\end{eqnarray}
}

A computation on an input $x$ of length $n$ corresponds to a unitary
evolution in the Hilbert space $\mathcal{H}_n = l_2(C_n)$. For each
$(q, a, b) \in C_n, q \in Q, a \in \mathbb{Z}, b \in [0, n + 1]$, let
$| q, a, b \rangle$ denote the basis vector in $l_2(C_n)$. An operator
$U^\delta_x$ for an input $x$ on $\mathcal{H}_n$ is defined as follows:
\begin{eqnarray*}
 U^\delta_x\ket{q, a, b} =
 \sum_{q^\prime, c, d}\delta(q, w_x(b), \sign(a), q^\prime, c, d)
  \ket{q^\prime, a + c, b + \mu(d)},
\end{eqnarray*}
where $w_x(b)$ is the $b$th symbol of $w_x = \cent x \dollar$ and
$\mu(d) = -1 (0) [+1]$ if $d = \leftarrow (\downarrow) [\rightarrow]$.
We assume that this operator is unitary, that is,
$\left(U_x^\delta\right)^* U_x^\delta = U_x^\delta
\left(U_x^\delta\right)^* = I$.

After each transition, a state of a 2Q1CA is observed. A computational 
observable $O$ corresponds to the orthogonal decomposition
$l_2(C_n) = \Eacc \oplus \Erej \oplus \Enon$. The outcome of any
observation will be either ``accept'' ($\Eacc$), ``reject'' ($\Erej$) or
``non-halting'' ($\Enon$). The probability of acceptance, rejection,
and non-halting at each step is equal to the sum of the squared
amplitude of each basis state in the new state for the corresponding subspace.
After the measurement, the state collapses to the projection to
one of $\Eacc, \Erej, \Enon$.

\begin{df}
 A language $L$ is {\bf\it recognizable} by a 2Q1CA with probability $p$,
 if there exists a 2Q1CA $M$ which accepts any input $x \in L$ with
 probability at least $p > 1/2$ and rejects any input $x \not\in L$ with
 probability at least $p$.
\end{df}

Finally we define two-way reversible one-counter automata,
which can be regarded as a special case of 2Q1CAs.

\begin{df}
 A two-way reversible one-counter automaton (2R1CA) is defined as a
 well-formed 2Q1CA whose transition amplitudes only take $0$ and $1$.
\end{df}

\begin{lem}
 A 2Q1CA satisfies the well-formedness conditions if and only 
 if $U_x^\delta$ is a unitary operator.
\end{lem}
\begin{proof}
 By the definition, $U_x^\delta$ transforms two quantum states
 $\ket{q_1, a_1, b_1}, \ket{q_2, a_2, b_2}$ into the following
 states:
\begin{eqnarray*}
 U^\delta_x \left| q_1, a_1, b_1 \right\rangle &=& 
  \sum_{q^\prime_1, c_1, d_1}
   \delta(q_1, \sigma_1, s_1, q^\prime_1, c_1, d_1)
    \ket{q^\prime_1, a_1 + c_1, b_1 + \mu(d_1)},\\
 U^\delta_x \left| q_2, a_2, b_2 \right\rangle &=&
  \sum_{q^\prime_2, c_2, d_2}
   \delta(q_2, \sigma_2, s_2, q^\prime_2, c_2, d_2)
    \ket{q^\prime_2, a_2 + c_2, b_2 + \mu(d_2)}.
\end{eqnarray*}

 First, we assume that $U_x^\delta$ is a unitary operator, that is,
 $\left(U_x^\delta\right)^* U_x^\delta = I$. Then we have the inner
 product of the previous two vectors as follows:
{\small
\begin{eqnarray*}
\lefteqn{
 \left\langle q_1, a_1, b_1 \left|\right.
              q_2, a_2, b_2 \right\rangle \,=\,
 \left\langle q_1, a_1, b_1 \left| \left(U^\delta_x\right)^* U^\delta_x
      \right| q_2, a_2, b_2 \right\rangle
    }\\
  &=&
 \sum_{q^\prime_1, c_1, d_1}\sum_{q^\prime_2, c_2, d_2}
  \delta^*(q_1, \sigma_1, s_1, q^\prime_1, c_1, d_1)
  \delta  (q_2, \sigma_2, s_2, q^\prime_2, c_2, d_2)
   \left\langle q^\prime_1, a_1 + c_1, b_1 + \mu(d_1) \left|\right.
                q^\prime_2, a_2 + c_2, b_2 + \mu(d_2) \right\rangle.
\end{eqnarray*}
}%
Here we have
\begin{eqnarray*}
\lefteqn{
 \left\langle q^\prime_1, a_1 + c_1, b_1 + \mu(d_1) \left|\right.
              q^\prime_2, a_2 + c_2, b_2 + \mu(d_2) \right\rangle
            }\nonumber\\
& = &
 \begin{cases}
  1 & q^\prime_1 = q^\prime_2,
      a_1 + c_1 = a_2 + c_2, b_1 + \mu(d_1) = b_2 + \mu(d_2), \\
  0 & \mbox{otherwise}.
 \end{cases}
\end{eqnarray*}
It follows that the inner product is
\begin{eqnarray}
 \left\langle q_1, a_1, b_1 \left|\right.
              q_2, a_2, b_2 \right\rangle =
 \sum_{{{q^\prime} \atop {a_1 + c_1 = a_2 + c_2}} \atop {b_1 + \mu(d_1) = b_2 + \mu(d_2)}}
  \delta^*(q_1, \sigma_1, s_1, q^\prime, c_1, d_1)
  \delta  (q_2, \sigma_2, s_2, q^\prime, c_2, d_2).
\end{eqnarray}

\begin{enumerate}
 \item In the case $a_1 = a_2$ and $b_1 = b_2$,
       left hand side of (14) = 1 if $q_1 = q_2$ and 0 otherwise. Thus $\delta$ 
       satisfies (1).
 \item In the case $a_1 = a_2$ and $b_1 = b_2 \pm 1$,
       left hand side of (14) = 0. Thus $\delta$ satisfies (2).
 \item In the case $a_1 = a_2$ and $b_1 = b_2 \pm 2$,
       left hand side of (14) = 0. Thus $\delta$ satisfies (3).
 \item In the case $a_1 = a_2 \pm 1$ and $b_1 = b_2$,
       left hand side of (14) = 0. Thus $\delta$ satisfies (4).
 \item In the case $a_1 = a_2 \pm 1$ and $b_1 = b_2 \pm 1$,
       left hand side of (14) = 0. Thus $\delta$ satisfies (5).
 \item In the case $a_1 = a_2 \pm 1$ and $b_1 = b_2 \mp 1$,
       left hand side of (14) = 0. Thus $\delta$ satisfies (6).
 \item In the case $a_1 = a_2 \pm 1$ and $b_1 = b_2 \pm 2$,
       left hand side of (14) = 0. Thus $\delta$ satisfies (7).
 \item In the case $a_1 = a_2 \pm 1$ and $b_1 = b_2 \mp 2$,
       left hand side of (14) = 0. Thus $\delta$ satisfies (8).
 \item In the case $a_1 = a_2 \pm 2$ and $b_1 = b_2$,
       left hand side of (14) = 0. Thus $\delta$ satisfies (9).
 \item In the case $a_1 = a_2 \pm 2$ and $b_1 = b_2 \pm 1$,
       left hand side of (14) = 0. Thus $\delta$ satisfies (10).
 \item In the case $a_1 = a_2 \pm 2$ and $b_1 = b_2 \mp 1$,
       left hand side of (14) = 0. Thus $\delta$ satisfies (11).
 \item In the case $a_1 = a_2 \pm 2$ and $b_1 = b_2 \pm 2$,
       left hand side of (14) = 0. Thus $\delta$ satisfies (12).
 \item In the case $a_1 = a_2 \pm 2$ and $b_1 = b_2 \mp 2$,
       left hand side of (14) = 0. Thus $\delta$ satisfies (13).
 \item In the case $|a_1 - a_2| > 2$ or $|b_1 - b_2| > 2$,
       left hand side of (14) = 0. Since
       two quantum states are always different from each other,
       right hand side of (14) = 0. 
       Thus, in this case, the equation (14) always holds.
\end{enumerate}
From these, we conclude that, if $U_x^\delta$ is a unitary operator 
$\delta$ satisfies the well-formedness conditions.

On the other hand, if we assume that $\delta$ satisfies the well-formedness
conditions, then we can easily check
$\left(U_x^\delta\right)^* U_x^\delta = I$, that is, $U_x^\delta$ is a
unitary operator.
\end{proof}

\subsection{Simple 2Q1CAs}
To describe automata easily, we introduce the concept of simple 2Q1CAs.
First we introduce {\it counter-simple 2Q1CAs}.

\begin{df}
 A 2Q1CA $(Q, \Sigma, \delta, q_0, \Qacc, \Qrej)$ is counter-simple,
 if there are linear operators $\{V_{\sigma, s}\}$ on $l_2(Q)$ and a
 counter function $C : Q \times \Gamma \rightarrow \{-1, 0, +1\}$ such
 that for any $q, q^\prime \in Q$, $\sigma \in \Gamma$, $s \in \{0, 1\}$,
 $c \in \{-1, 0, +1\}$, $d \in \{\leftarrow, \downarrow, \rightarrow\}$,
 \begin{eqnarray*}
  \delta(q, \sigma, s, q^\prime, c, d) =
   \begin{cases}
    \bra{q^\prime}V_{\sigma, s}\ket{q} & C(q^\prime, \sigma) = c,\\
    0 & \mbox{\rm otherwise},
   \end{cases}
 \end{eqnarray*}
 where $\bra{q^\prime}V_{\sigma, s}\ket{q}$ is the coefficient of
 $\ket{q^\prime}$ in $V_{\sigma, s}\ket{q}$.
\end{df}
In this case, increase or decrease of the counter value is
determined by the new state and the symbol it reads. Thus the 2Q1CA
satisfies the conditions (4), (5), (9) automatically.

Next we introduce {\it head-simple 2Q1CAs}.

\begin{df}
 A 2Q1CA $(Q, \Sigma, \delta, q_0, \Qacc, \Qrej)$ is head-simple,
 if there are linear operators $\{V_{\sigma, s}\}$ on $l_2(Q)$ and a
 head function $D : Q \rightarrow \{\leftarrow, \downarrow, \rightarrow\}$ such
 that for any $q, q^\prime \in Q$, $\sigma \in \Gamma$, $s \in \{0, 1\}$,
 $c \in \{-1, 0, +1\}$, $d \in \{\leftarrow, \downarrow, \rightarrow\}$,
 \begin{eqnarray*}
  \delta(q, \sigma, s, q^\prime, c, d) =
   \begin{cases}
    \bra{q^\prime}V_{\sigma, s}\ket{q} & D(q^\prime) = d,\\
    0 & \mbox{\rm otherwise}.
   \end{cases}
 \end{eqnarray*}
\end{df}
In this case, the tape head move is determined by the new state. Thus the
2Q1CA satisfies the conditions (2), (3), (5), (6), (7), (8), (10), (11), 
(12), (13) automatically.

Finally we introduce {\it simple 2Q1CAs} combining the concepts of
counter-simple 2Q1CAs and head-simple 2Q1CAs.

\begin{df}
 A 2Q1CA $(Q, \Sigma, \delta, q_0, \Qacc, \Qrej)$ is simple,
 if there are linear operators $\{V_{\sigma, s}\}$ on $l_2(Q)$, a
 counter function $C : Q \times \Gamma \rightarrow \{-1, 0, +1\}$ and a
 head function $D : Q \rightarrow \{\leftarrow, \downarrow, \rightarrow\}$ such
 that for any $q, q^\prime \in Q$, $\sigma \in \Gamma$, $s \in \{0, 1\}$,
 $c \in \{-1, 0, +1\}$, $d \in \{\leftarrow, \downarrow, \rightarrow\}$,
 \begin{eqnarray*}
  \delta(q, \sigma, s, q^\prime, c, d) =
   \begin{cases}
    \bra{q^\prime}V_{\sigma, s}\ket{q} & {C(q^\prime, \sigma) = c, D(q^\prime) = d},\\
    0 & \mbox{\rm otherwise}.
   \end{cases}
 \end{eqnarray*}
\end{df}
In this case, increase or decrease of the counter value is
determined by the new state and the symbol it reads, and the head move is
determined by the new state. Thus the 2Q1CA satisfies the conditions
(2)--(13) automatically.

Notice that the transition function of 2D1CAs can also be written as
$\delta : Q \times \Gamma \times S \times
Q \times \{-1, 0, +1\} \times \{\leftarrow, \downarrow, \rightarrow\}
\rightarrow \{0, 1\}$, that is,
\begin{eqnarray*}
 \delta(q, \sigma, s) = (q^\prime, c, d) &\Leftrightarrow&
 \delta(q, \sigma, s, q^\prime, c^\prime, d^\prime) =
 \begin{cases}
  1 & (c^\prime, d^\prime) =   (c, d),\\
  0 & (c^\prime, d^\prime) \ne (c, d).
 \end{cases}
\end{eqnarray*}
Thus we can also define simple 2D1CAs in the way similar to simple
2Q1CAs.

\begin{lem}
 A simple 2Q1CA satisfies the well-formedness conditions if
 there are unitary operators $\{V_{\sigma, s}\}$ such that
 for any $\sigma \in \Gamma$ and $s \in \{0, 1\}$,
 \begin{eqnarray}
  \sum_{q^\prime}\bra{q^\prime}V_{\sigma, s}\ket{q_1}^*
                 \bra{q^\prime}V_{\sigma, s}\ket{q_2} =
   \begin{cases}
    1 & q_1 =  q_2,\\
    0 & q_1 \ne q_2.
   \end{cases}
 \end{eqnarray}
\end{lem}
\begin{proof}
 By the definition of simple 2Q1CAs, the well-formedness conditions
 except for (1) are satisfied automatically.

 Now, we let
 \begin{eqnarray*}
  \delta(q_1, \sigma, s, q^\prime, c, d) &=&
   \begin{cases}
    \bra{q^\prime}V_{\sigma, s}\ket{q_1} & {c = C(q^\prime, \sigma), d = D(q^\prime)},\\
    0 & \mbox{otherwise},
   \end{cases}\\
  \delta(q_2, \sigma, s, q^\prime, c, d) &=&
   \begin{cases}
    \bra{q^\prime}V_{\sigma, s}\ket{q_2} & {c = C(q^\prime, \sigma), d = D(q^\prime)},\\
    0 & \mbox{otherwise},
   \end{cases}
 \end{eqnarray*}
 then it is trivial to show that a simple 2Q1CA satisfies (1) if
 it satisfies (15).
\end{proof}

\section{Reversible Simulation of 2D1CAs}
In this section, we show that an arbitrary 2D1CA is simulated by a
2R1CA (which is viewed as a special case of 2Q1CA by its definition).
To make the reversible simulation simpler,
we show the following lemma.

\begin{lem}
 For any language $L$ which is recognized by a 2D1CA, there exists a
 simple 2D1CA which recognizes $L$ and whose counter function does not
 depend on the symbol it reads.\label{simp}
\end{lem}
\begin{proof}
 Let $M = (Q, \Sigma, \delta, q_0, \Qacc, \Qrej)$ be a 2D1CA which
 recognizes $L$. Consider that $M$ is in the state $q^\prime$ and
 there are transitions such that $\delta(q_1, \sigma_1, s_1) =
 (q^\prime, c_1, d_1), \delta(q_2, \sigma_2, s_2) = (q^\prime, c_2, d_2),
 (c_1, d_1) \ne (c_2, d_2)$. Since $M$ is a 2D1CA, its previous state
 cannot be a superposition of $q_1$ and $q_2$. This means that we can
 distinguish $q^\prime$ reached by the transition $\delta(q_1, \sigma_1, s_1) =
 (q^\prime, c_1, d_1)$ from $q^\prime$ reached by the transition
 $\delta(q_2, \sigma_2, s_2) = (q^\prime, c_2, d_2)$.

We construct a simple 2D1CA for $L$ as follows.
 Let the state sets
 \begin{eqnarray*}
  Q^\prime &=& \left\{q^{c, d} \mid q \in Q,
  c \in \{-1, 0, +1\}, d \in \{\leftarrow, \downarrow, \rightarrow\}\right\},\\
  \Qacc^\prime &=& \left\{q^{c, d} \mid q \in \Qacc,
  c \in \{-1, 0, +1\}, d \in \{\leftarrow, \downarrow, \rightarrow\}\right\},\\
  \Qrej^\prime &=& \left\{q^{c, d} \mid q \in \Qrej,
  c \in \{-1, 0, +1\}, d \in \{\leftarrow, \downarrow, \rightarrow\}\right\}.
 \end{eqnarray*}
For each $q_1^{c_1, d_1}, q_2^{c_2, d_2} \in Q^\prime, \sigma \in \Gamma$,
 we define transition matrices $\{V^{\prime}_{\sigma, s}\}$, a counter function
 $C^{\prime}$, and a head function $D^{\prime}$ as follows:
 \begin{eqnarray*}
  V^{\prime}_{\sigma, s}\ket{q_1^{c_1, d_1}} &=& q_2^{c_2, d_2}
  ~~~~~\mbox{for}~ \delta(q_1, \sigma, s) = (q_2, c_2, d_2),\\
  C^{\prime}(q^{c, d}, \sigma) &=& c \hspace{1.3cm} \mbox{for any}~ \sigma \in \Gamma,\\
  D^{\prime}(q^{c, d}) &=& d,
 \end{eqnarray*}
 and let the transition function $\delta^\prime$ of $M^\prime$
be defined from given $V^{\prime}_{\sigma, s}, C^{\prime}$, and $D^{\prime}$. 
 Then $M^\prime = (Q^\prime, \Sigma, \delta^\prime,
 q_0^{0, \downarrow}, \Qacc^\prime, \Qrej^\prime)$ is a simple 2D1CA and
 by its construction we can easily check that $M^\prime$ simulates $M$
 and the counter function $C^\prime$ of $M^\prime$ does not depend on the symbol it
 reads.
\end{proof}

Notice that simple 2D1CAs cannot be more powerful than 2D1CAs by their
definition. Therefore we have the following result.
\begin{cor}
 Simple 2D1CAs are as powerful as 2D1CAs.
\end{cor}

Now we prove that an arbitrary 2D1CA is simulated by a simple 2R1CA
(and hence by a simple 2Q1CA) by using the technique of reversible simulation.
Kondacs and Watrous \cite{KondacsW1997} showed that
any two-way deterministic finite state automata
can be simulated by two-way reversible finite state automata
by using a technique due to Lange, McKenzie, and Tapp \cite{LangeMT1997}.
Our proof is an extension of the proof by
Kondacs and Watrous \cite{KondacsW1997} to one-counter cases.

\begin{Th}
 For any language $L$ which is recognized by a 2D1CA, there exists a
 simple 2R1CA which recognizes $L$.
\end{Th}
\begin{proof}
 Let $M = (Q, \Sigma, \delta, q_0, \Qacc, \Qrej)$ be a 2D1CA which
 recognizes $L$.
 From Lemma \ref{simp} we assume that $M$ is a simple
 2D1CA and its counter function does not depend on the symbol it
 reads. Given such an $M$, we define a 2R1CA $M^\prime =
 \bigl(Q^\prime, \Sigma, \delta^\prime, q_0^\prime, \Qacc^\prime,
 \Qrej^\prime\bigr)$ which recognizes $L$.

 First, define
 \begin{eqnarray*}
  I_{q, \sigma, s} &=&
   \left\{q^\prime \in Q \mid \delta(q^\prime, \sigma, s) = \delta(q, \sigma, s)\right\},\\
  J_{q, \sigma, s} &=&
   \left\{q^\prime \in Q \mid \delta(q^\prime, \sigma, s) = q\right\},
 \end{eqnarray*}
 and fix an ordering of the set $Q$. Let $\max(\cdot)$ and $\min(\cdot)$
 denote the maximum and minimum functions relative to this ordering, and 
 for any subset $R \subset Q^\prime$, let $\mathrm{succ}(q, R)$ be the
 least element larger than $q \in R$.

 Now we define $M^\prime$. Let the state sets $Q^\prime = Q \times \{-, +\}$,
 $\Qacc^\prime = \Qacc \times \{-, +\}$,
 $\Qrej^\prime = \Qacc \times \{-, +\}$, and let $q_0^\prime = (q_0, +)$.
 Also define the transition matrices $\{V^\prime_{\sigma, s}\}$, the
 counter function $C^{\prime}$, and the head function $D^{\prime}$
 of $M^{\prime}$ as follows.

 For each $q \in Q, \sigma \in \Gamma$ for which $V_{\sigma, s}$
 is defined in $M$,
 \begin{eqnarray*}
  V^\prime_{\sigma, s}\ket{(q, +)} =
   \begin{cases}
    \ket{\bigl(
          \mathrm{succ}(q, I_{q, \sigma, s}), -
         \bigr)} & q \ne \max(I_{q, \sigma, s}),\\
    \ket{(V_{\sigma, s}\ket{q}, +)} & q = \max(I_{q, \sigma, s}).
   \end{cases}
 \end{eqnarray*}
For each $q \in Q^\prime, \sigma \in \Gamma$,
 \begin{eqnarray*}
  V^\prime_{\sigma, s}\ket{q, -} =
   \begin{cases}
    \ket{(q, +)}                   & \hspace{7mm}J_{q, \sigma, s} = \emptyset,\\
    \ket{\bigl(
          \min(J_{q, \sigma, s}), -
         \bigr)} & \hspace{7mm}J_{q, \sigma, s} \ne \emptyset.
   \end{cases}
 \end{eqnarray*}
 Define $C^\prime\bigl((q, +), \sigma\bigr) = C(q, \sigma),
 C^\prime\bigl((q, -), \sigma\bigr) = -C(q, \sigma)$, and
 $D^\prime\bigl((q, +)\bigr) = D(q), D^\prime\bigl((q, -)\bigr) = -D(q)$.
 Let the transition function $\delta^\prime$ of $M^\prime$
be defined from given $V^{\prime}_{\sigma, s}, C^{\prime}$, and $D^{\prime}$.

 For given $M$ and any input $x$ of length $n$, let $G$ be an undirected 
 graph with set of vertices $Q^\prime \times \mathbb{Z} \times [0, n + 1]$,
 and edges between vertices $(q_1, a, b)$ and $\bigl(q_2, a + C(q_2, \sigma),
 b + D(q_2)\bigr)$ if and only if $V_{w_x(b), \sign(a)}\ket{q_1} = q_2$.
 Let $G_0$ be the connected component of $G$ which contains the initial
 configuration $\bigl((q_0, +), 0, 0\bigr)$. There can be no cycles in
 $G_0$, and $G_0$ must contain exactly one vertex corresponding to a
 halting state $\in \Qacc \cup \Qrej$. Thus we can view $G_0$ as a tree
 with the single halting configuration vertex as the root. $M^\prime$
 simulates $M$ by traversing $G_0$ in a reversible manner.

 For each configuration $(q, a, b)$ of $M$, there correspond two
 configurations $\bigl((q, +), a, b\bigr)$ and
 $\bigl((q, -), a - C(q, \sigma), b - \mu(D(q))\bigr)$ of $M^\prime$, which
 are to be interpreted as follows. Configuration
 $\bigl((q, +), a, b\bigr)$ indicates that the subtree of $G_0$ rooted
 at vertex $(q, a, b)$ has just been traversed, and configuration
 $\bigl((q, -), a - C(q, \sigma), b - \mu(D(q))\bigr)$ indicates that the
 subtree of $G_0$ rooted at vertex $(q, a, b)$ is now about to be
 traversed (here notice that $C$ does not depend on $\sigma$).

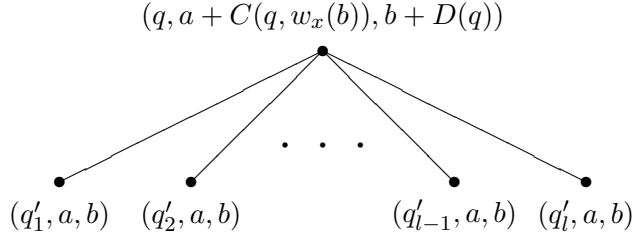
\begin{figure}[t]
\begin{center}
\setlength{\unitlength}{1mm}
\begin{picture}(110,32)
\put(12.5,0){\makebox(15,5){$(q^{\prime}_{1}, a, b)$}}
\put(30,0){\makebox(15,5){$(q^{\prime}_{2}, a, b)$}}
\put(65,0){\makebox(15,5){$(q^{\prime}_{l-1}, a, b)$}}
\put(82.5,0){\makebox(15,5){$(q^{\prime}_{l}, a, b)$}}
\put(40,26.5){\makebox(30,5){$(q, a+C(q,w_x(b)), b+D(q))$}}
\put(55,24.5){\line(-2,-1){35}}
\put(55,24.5){\line(-1,-1){17.5}}
\put(55,24.5){\line(1,-1){17.5}}
\put(55,24.5){\line(2,-1){35}}
\put(55,24.5){\circle*{1.5}}
\put(20,7){\circle*{1.5}}
\put(37.5,7){\circle*{1.5}}
\put(72.5,7){\circle*{1.5}}
\put(90,7){\circle*{1.5}}
\multiput(50,12)(5,0){3}{\circle*{0.75}}
\end{picture}
\caption{Vertex $(q, a+C(q,w_x(b)), b+\mu(D(q)))$ and its children}
\label{subtree}
\end{center}
\end{figure}

 Consider Figure \ref{subtree}. Here we have $J_{q, \sigma, s} =
 I_{q^\prime, \sigma, s} = \{q_1^\prime, \dots, q_l^\prime\}$ for
 each $i = 1, \dots, l$, where $\sigma = w_x(b)$ and $s = \sign(a)$.
 We assume that $q_1^\prime < \dots < q_l^\prime$ according to our
 ordering of $Q^\prime$. Suppose that $M^\prime$ is in configuration
 $\bigl((q_i^\prime, +), a, b\bigr)$ for $i < l$. Since
 $q_i^\prime \ne \max(I_{q_i^\prime, \sigma, s})$, the next
 configuration of $M^\prime$ is
 \begin{eqnarray*}
  & &\hspace*{-1cm}
  \Bigl(
   \bigl(
    \mathrm{succ}(q_i^\prime, I_{q_i^\prime, \sigma, s}), -
   \bigr),
   a + C^\prime\bigl((
                \mathrm{succ}(q_i^\prime, I_{q_i^\prime, \sigma, s}), -
               ), \sigma\bigr),
   b + \mu\bigl(D^\prime\bigl((
           \mathrm{succ}(q_i^\prime, I_{q_i^\prime, \sigma, s}), -
          )\bigr)\bigr)
  \Bigr)\\
  &=& \!\!
  \Bigl(
   \bigl(
    q_{i + 1}^\prime, -
   \bigr),
   a - C(q_i^\prime, \sigma),
   b - \mu(D(q_i^\prime))
  \Bigr).
 \end{eqnarray*}
 Now suppose that $M^\prime$ is in configuration
 $\bigl((q_l^\prime, +), a, b\bigr)$. Since
 $q_i^\prime = \max(I_{q_i^\prime, \sigma, s})$, the next
 configuration is
 {\small
 \begin{eqnarray*}
  \Bigl(
   \bigl(
    V_{\sigma, s}\ket{q_l^\prime}, +
   \bigr),
   a + C^\prime\bigl((
                V_{\sigma, s}\ket{q_l^\prime}, +
               ), \sigma\bigr),
   b + \mu\bigl(D^\prime\bigl((
           V_{\sigma, s}\ket{q_l^\prime}, +
          )\bigr)\bigr)
  \Bigr) =
  \Bigl(
   \bigl(
    q, +
   \bigr),
   a + C(q, \sigma),
   b + \mu(D(q))
  \Bigr).
 \end{eqnarray*}
 }%
 Hence, $M^\prime$ enters configuration $\bigl((q, +), a + C(q, \sigma),
 b + \mu(D(q))\bigr)$ only after each of the subtrees rooted at its
 children have been traversed. Next, suppose that $M^\prime$ is in
 configuration $\left((q, -), a - C(q, \sigma), b - D(q)\right)$. The
 next configuration of $M^\prime$ is
 \begin{eqnarray*}
  & &\hspace*{-1.5cm}
  \Bigl(
   \bigl(
    \min(J_{q, \sigma, s}), -
   \bigr),
   a - C(q, \sigma) + C^\prime\bigl((
                               \min(J_{q, \sigma, s}), -
                              )\bigr),
   b - \mu(D(q)) + \mu\bigl(D^\prime\bigl((
                       \min(J_{q, \sigma, s}), -
                      )\bigr)\bigr)
  \Bigr)\\
  &=& \!\!
  \Bigl(
   \bigl(
    q_1^\prime, -
   \bigr),
   a - C(q, \sigma) - C(q_1^\prime, \sigma),
   b - \mu(D(q)) - \mu(D(q_1^\prime))
  \Bigr),
 \end{eqnarray*}
 and so the subtree rooted at vertex $(q_1^\prime, a - C(q, \sigma),
 b - \mu(D(q)))$ is now to be traversed.
 Finally, in the case that $\bigl(q, a + C(q, \sigma), b + \mu(D(q))\bigr)$
 has no predecessors, we have $J_{q, \sigma, s} = \emptyset$, and so the 
 configuration which immediately follows $\bigl((q, -), a, b\bigr)$ is
 $\bigl((q, +), a + C(q, \sigma), b + \mu(D(q))\bigr)$.

 By traversing $G_0$ in this manner, $M^\prime$ eventually enters one of
 the configurations in $\Qacc^\prime \times \mathbb{Z} \times [0, n + 1]$
 or $\Qrej^\prime \times \mathbb{Z} \times [0, n + 1]$, and consequently
 accepts or rejects accordingly. It is clear that $M^\prime$ recognizes
 $L$ since $M$ recognizes $L$.
\end{proof}

\section{Recognizability}
\subsection{2Q1CA for $\boldsymbol{\{a^n b^{n^2} \mid n \ge 1\}}$}\label{anbn2}
\begin{prop}
 Let $L_{\rm square}$ be $\{a^n b^{n^2} \mid n \ge 1\}$.
 For an arbitrary fixed positive integer $N \ge 2$,
 there exists a 2Q1CA $M_{\rm square}$which
 accepts $x \in L_{\rm square}$ with probability $1$ and
 rejects $x \not\in L_{\rm square}$ with probability
 $1 - 1/N$. In either case, $M_{\rm square}$ halts after $O(N|x|)$ steps
 with certainty.
\end{prop}
\begin{proof}
 Let $Q = \{q_0, q_1, q_2, q_3, q_4, q^i_{5, j_1},
 q^i_{6, j_2}, q^i_7 \mid 1 \le i \le N, 1 \le j_1 \le i,
 1 \le j_2 \le N - i + 1\}$, $\Qacc = \{q^N_7\}$ and
 $\Qrej = \{q^j_7 \mid 1 \le j \le N - 1\}$. For each $q \in Q$,
 $\sigma \in \Gamma$, $s \in \{0, 1\}$, we define the transition
 matrices $\{V_{\sigma, s}\}$, the counter function $C$, and the head
 function $D$ as follows:
 {\small
\begin{eqnarray*}
 \begin{array}{lll}
  \begin{array}{l}
   \trans{\cent, 0}{0}{0}\\
   \Trans{\cent, 0}{5, 0}{6, N - i + 1}\\
   \\
   \trans{\dollar, 0}{2}{3}\\
   V_{\dollar, 0}\ket{q^{i}_{6, N - i + 1}} = \frac{1}{\sqrt{N}}\sum_{k = 1}^Ne^{\frac{2\pi i k}{N}\sqrt{-1}}\ket{q^k_7}\\
   \\
   \begin{array}{@{}ll}
    \begin{array}{@{}l}
     C(q^i_{5, 2i}, a) = -1\\
     C(q^i_{5,  i}, a) = +1\\
     C(q^i_{5, 2i}, b) = -1\\
     C(q^i_{5,  0}, a) = +1\\
     C(q, \sigma)      =  0\\
      \quad \mbox{otherwise}
    \end{array} &
    \begin{array}{l}
     D(q_j)         = \rightarrow ~~ j = 0, 2, 4\\
     D(q_j)         = \leftarrow  ~~ j = 1, 3\\
     D(q^i_{5, 2i}) = \leftarrow\\
     D(q^i_{5,  i}) = \rightarrow\\
     D(q^{\i}_{6, N - i + 1}) = \rightarrow\\
     D(q) = \downarrow ~~ \mbox{otherwise}\\
    \end{array}
   \end{array}
  \end{array} &
  \begin{array}{l}
   \trans{a, 0}{0}{0}\\
   \trans{a, 0}{1}{2}\\
   \trans{a, 0}{3}{4}\\
   \Trans{a, \*}{5, j + 1}{5, j}\\
    \qquad j \ne 0, i\\
   \Trans{a, 0}{5, i + 1}{5,  i}\\
   \Trans{a, 1}{5, i + 1}{5, 2i}\\
   \Trans{a, 1}{5,     1}{5,  i}\\
   \Trans{a, 0}{6, j + 1}{6,  j}\\
    \qquad 1 \le j \le N - i\\
   \Trans{a, 0}{6, 1}{6, N - i + 1}\\
  \end{array} &
  \begin{array}{l}
   \trans{b, 0}{0}{1}\\
   \trans{b, 0}{2}{2}\\
   \trans{b, 0}{3}{3}\\
   V_{b, 0}\ket{q_4} = \frac{1}{\sqrt{N}}\sum_{i = 1}^N\ket{q^i_{5, 0}}\\
   \Trans{b, 1}{5, i}{5, 0}\\
   \Trans{b, 1}{5, 0}{5, 2i}\\
   \Trans{b, 0}{6, j + 1}{6, j}\\
   \Trans{b, 0}{6, 1}{6, N - i + 1}\\
   \\
   \\
   \\
  \end{array}
 \end{array}
\end{eqnarray*}
 }

\begin{figure}[t]
 \begin{center}
  \includegraphics[width=5cm]{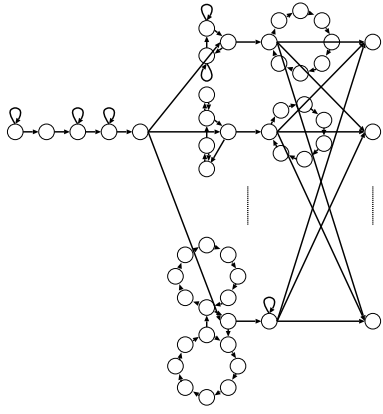}
 \end{center}
 \caption{Transition image of $M_{\rm square}$}
\end{figure}

 By the construction of $M_{\rm square}$, we see that the computation
 consists of three phases. The first phase rejects any input not
 of the form $a^+ b^+$. This phase is straightforward, similar to
 2-way reversible finite
 automata (with no-counter) which recognizes the input of the form
 $a^+ b^+$. If the input is not of the form $a^+ b^+$, the computation 
 terminates and rejects.
 Otherwise, the second phase begins with the state $q_4$
 with the tape head reading the left-most $b$.

 At the start of the second phase, the computation branches into $N$
 paths, indicated by the states $q^1_{5, 0}, \dots, q^N_{5, 0}$, each
 with amplitude $1/\sqrt{N}$. For each of these paths, $M_{\rm square}$
 moves the tape head to left and
 right deterministically in the following way.
 Along the $i$th path,
 while the counter value is not 0,
 the automaton decreases the counter value by 1 and moves the tape head to
 left.
 In addition, every time the tape head reads the symbol $a$,
 it remains stationary for $i$ steps.
 Upon the counter value being 0,
 it repeats the followings
 until the tape head reads the left-most $b$:
 it increases the counter value by 1 and moves the tape 
 head to right. 
 If the tape
 head reads the symbol $\cent$,
 the computation enters the third phase with the state
 $q^i_{5, 0}$.
 Thus, while $M_{\rm square}$ is scanning $a$'s in the
 input during the second phase, the tape head requires precisely
 \begin{eqnarray*}
  (i + 1)\left(\sum_{c = 0}^{m - 1}(2c + 1) + m\right)
   = (i + 1)\left(m^2 + m\right)
 \end{eqnarray*}
 steps along the $i$th path, where $m$ is the number of $a$'s.

 Along the $i$th path on the third phase, if the tape head reads the
 symbol $a$ or $b$ then it remains stationary for $N - i + 1$ steps and
 after that moves to right. Upon reading the symbol $\dollar$, each
 computation path again splits according to the Discrete-Fourier
 Transformation, yielding the single accepting state $q^N_7$ and the
 other rejecting states $q^i_7 (1 \le i \le N - 1)$. Thus, while the
 automaton is scanning $a$'s and $b$'s in the input, the tape head
 requires precisely $(N - i + 1)\left(m + n\right)$ steps along the
 $i$th path, where $n$ is the number of $b$'s.
 Therefore, it is easy to see that, under the assumption
 $i \ne i^\prime$,
 $(i + 1)\left(m^2 + m\right) + (N - i + 1)\left(m + n\right) =
  (i^\prime + 1)\left(m^2 + m\right) + (N - i^\prime + 1)\left(m + n\right)$
 if and only if $m^2 = n$.
 
 First consider the case that $m^2 = n$. Since each of the $N$
 computation paths reaches the symbol $\dollar$ at the same time,
 the superposition immediately after performing the
 Discrete-Fourier Transformation is
 \begin{eqnarray*}
  \frac{1}{N}\sum_{i = 1}^N\sum_{k = 1}^Ne^{\frac{2\pi ik}{N}\sqrt{-1}}\ket{q^k_7} = \ket{q^N_7}.
 \end{eqnarray*}
 Hence, the accepting state $q^N_7$ is entered with probability 1.

 Next suppose that $m^2 \ne n$. In this case, each of $N$
 computation paths reaches the symbol $\dollar$ at a different timing.
 Thus, there is no cancellation among the rejection states. For
 each of $N$ paths, the conditional probability that an observation
 results in $q^N_7$ at the time is $1/N$. It follows that the total
 probability that an observation results in $q^N_7$ is also
 $1/N$. Consequently the input is rejected with probability $1 - 1/N$.

 We clearly see that each possible computation path has length
 $O(N|x|)$, since each path terminates in a halting states with certainty.
\end{proof}

\subsection{2Q1CA for $\boldsymbol{\{a^m b^n c^{mn} \mid m, n \ge 1\}}$}
\begin{prop}\label{seeappendix}
 Let $L_{\rm prod}$ be $\{a^m b^n c^{mn} \mid m, n \ge 1\}$.
 For an arbitrary fixed positive integer $N \ge 2$, there exists a
 2Q1CA $M_{\rm prod}$ which
 accepts $x \in L_{\rm prod}$ with probability $1$ and
 rejects $x \not\in L_{\rm prod}$ with probability
 $1 - 1/N$. In either case, $M_{\rm prod}$ halts after $O(N|x|)$ steps
 with certainty.
\end{prop}
\begin{proof}
 Let $Q = \{q_0, q_1, q_2, q_3, q_4, q_5, q^i_6, q^i_7,
 q^i_{8, j_1}, q^i_9, q^i_{10, j_2}, q^i_{11}
 \mid 1 \le i \le N, 1 \le j_1, \le i, 1 \le j_2 \le N - i + 1\}$,
 $\Qacc = \{q^N_{11}\}$ and $\Qrej = \{q^j_{11} \mid 1 \le j \le N - 1\}$.
 For each $q \in Q$,
 $\sigma \in \Gamma$, $s \in \{0, 1\}$, we define the transition
 matrices $\{V_{\sigma, s}\}$, the counter function $C$ and the head
 function $D$ as follows:

{\small
\begin{eqnarray*}
 \begin{array}{@{}l@{}l@{}l@{}l}
  \begin{array}{l}
   \trans{\cent, 0}{0}{0}\\
   V_{\cent, 0}\ket{q_5} = \frac{1}{\sqrt{N}}\sum_{i = 1}^N\ket{q^i_6}\\
   \Trans{\cent, 1}{9}{6}\\
   \\
   \trans{\dollar, 0}{4}{5}\\
   V_{\dollar, 0}\ket{q^i_{10, 1}} = \frac{1}{\sqrt{N}}\sum_{k = 1}^Ne^{\frac{2\pi ik}{N}\sqrt{-1}}\ket{q^k_{11}}\\
   \\
   C(q^i_6, a) = +1 \hfil \qquad D(q_j) = \rightarrow\\
   C(q^i_9, a) = -1 \hfil \qquad  \qquad j = 0, 2, 4\\
   C(q^i_9, c) = -1 \hfil \qquad D(q_j) = \leftarrow\\
   C(q, \sigma) = 0     \hfil \qquad  \qquad j = 1, 3, 5\\
    \qquad \mbox{otherwise} \hfil \qquad \hspace{-.5mm} D(q^i_6) = \rightarrow\\
   \\
  \end{array} &
  \begin{array}{l}
   \trans{a, 0}{0}{0}\\
   \trans{a, 0}{1}{2}\\
   \trans{a, 0}{5}{5}\\
   \Trans{a, \*}{6}{6}\\
   \Trans{a, \*}{9}{9}\\
   \Trans{a, 1}{7}{8, 1}\\
   \\
   D(q^i_7) = \leftarrow\\
   D(q^i_{8, 1}) = \rightarrow\\
   D(q^i_9) = \leftarrow\\
   D(q^i_{10}) = \rightarrow\\
   D(q) = \downarrow\\
    \qquad \mbox{otherwise}\\
  \end{array} &
  \begin{array}{l}
   \trans{b, 0}{0}{1}\\
   \trans{b, 0}{2}{2}\\
   \trans{b, 0}{3}{4}\\
   \trans{b, 0}{5}{5}\\
   \Trans{b, 0}{9}{10, 1}\\
   \Trans{b, 1}{6}{7}\\
   \Trans{b, 1}{8, j}{8, j + 1}\\
    \qquad 1 \le j \le i - 1\\
   \Trans{b, 1}{8, i}{8, 1}\\
   \Trans{b, 1}{9}{9}\\
   \\
   \\
   \\
  \end{array} &
  \begin{array}{l}
   \trans{c, 0}{2}{3}\\
   \trans{c, 0}{4}{4}\\
   \trans{c, 0}{5}{5}\\
   \Trans{c, 0}{10, j}{10, j + 1}\\
    \qquad 1 \le j \le N - i\\
   \Trans{c, 0}{10, N - i + 1}{10, 1}\\
   \Trans{c, 1}{8, 1}{9}\\
   \\
   \\
   \\
   \\
   \\
   \\
  \end{array}
 \end{array}
\end{eqnarray*}
}

\begin{figure}[t]
 \begin{center}
 \includegraphics[width=8cm]{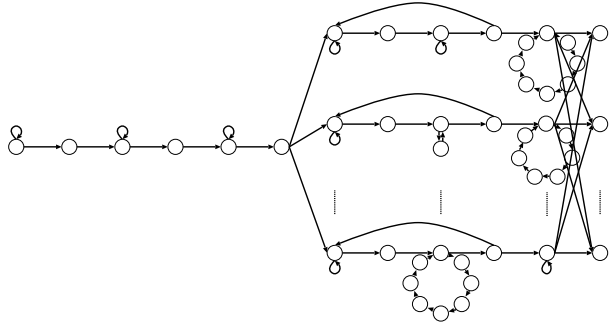}
 \end{center}
 \caption{Transition image of $M_{\rm prod}$}
\end{figure}

 By the construction of the automaton, we see that the computation
 consists of four phases. The first phase rejects any input not of the
 form $a^+ b^+ c^+$. This phase is straightforward, similar to 2-way
 reversible finite
 automata (with no-counter) which recognizes the input of the form
 $a^+ b^+ c^+$. If the input is not of the indicated form, the computation 
 terminates. Otherwise, the second phase begins with the state $q_5$
 with the tape head reading the symbol $\cent$.

 At the start of the second phase, the computation branches into $N$
 paths, indicated by the states $q^1_6, \dots, q^N_6$, each
 with amplitude $1/\sqrt{N}$.
 For each of these paths, the automaton moves the tape head to right and 
 increases the counter value by 1 while reading the symbol $a$. Upon
 reading the symbol $b$, the third phase begins with the state
 $q^i_8$.

 Along the $i$th path on the third phase,
 if the tape head reads the
 symbol $b$ then it remains stationary for $i$ steps and after that
 moves to right.
 After reading the symbol $c$,
 if the counter value is 0,
 the fourth phase begins with the state $q^i_{10}$.
 Otherwise repeat the followings
 until the tape head reaches the symbol $\cent$.
 While reading the symbol $b$,
 the automaton moves the tape head to left.
 While reading the symbol $a$,
 the automaton moves the tape head to left
 and decreases the counter value by 1.
 Upon reading the symbol $\cent$,
 the automaton re-enters the second phase.
 Thus, the tape head requires precisely $m$ steps on each second phase 
 and $i n + m + n$ steps along the $i$th path on each third phase. It is
 easy to see that the automaton repeats the second and the third
 phase $m$ times.

 Along the $i$th path on the fourth phase, if the tape head reads the
 symbol $c$ then it remains stationary for $N - i + 1$ steps and after
 that moves to right. Upon reading the symbol $\dollar$, each
 computation path again splits according to the Discrete-Fourier
 Transformation, yielding the single accepting state $q^N_{11}$ and the 
 other rejecting states $q^i_{11} (1 \le i \le N - 1)$. Thus, the
 tape head requires precisely $(N - i + 1)l$ steps along the $i$th
 path, where $l$ is the number of $c$'s.

 Therefore, it is easy to see that, under assumption $i \ne i^\prime$,
 $2m^2 + (i+1) mn + (N - i + 1)l =
 2m^2 + (i^\prime + 1) mn + (N - i^\prime + 1)l$ if and only if $l = mn$.

 First consider the case that $l = mn$. Since each of the $N$
 computation paths reaches the symbol $\dollar$ at the same time,
 the superposition immediately after performing the
 Discrete-Fourier Transformation is
 \begin{eqnarray*}
  \frac{1}{N}\sum_{i = 1}^N\sum_{k = 1}^Ne^{\frac{2\pi ik}{N}\sqrt{-1}}\ket{q^k_{11}} = \ket{q^N_{11}}.
 \end{eqnarray*}
 Hence, the accepting state $q^N_{11}$ is entered with probability 1.

 Next suppose that $l \ne mn$. In this case, each of $N$
 computation paths reaches the symbol $\dollar$ at a different timing.
 Thus, there is no cancellation among the rejection states. For
 each of $N$ paths, the conditional probability that an observation
 results in $q^N_{11}$ at the time is $1/N$. It follows that the total
 probability that an observation results in $q^N_{11}$ is also
 $1/N$. Consequently the input is rejected with probability $1 - 1/N$.

 We clearly see that each possible computation path has length
 $O(N|x|)$, since each path terminates in a halting states with
 certainty.
\end{proof}

\subsection{2Q1CA for
 $\boldsymbol{\{a_1^{n^1} a_2^{n^2} \dots a_k^{n^k} \mid n \ge 1\}}$}
\begin{prop}
 For each fixed $k \ge 1$, there exists a 2Q1CA which recognizes
 $\{a_1^{n^1} a_2^{n^2} \dots a_k^{n^k} \mid n \ge 1\}$.
\end{prop}
\begin{sketchofproof}
 In the case $k = 1$, it is trivial. In the case $k = 2$, we
 proved in Proposition 13.
 Consider the case $k \ge 3$.
 First, the automaton checks whether the input is of the form
 $a_1^+ a_2^+ \dots a_k^+$. Next, the automaton checks whether
 $m_1^2 = m_2$ or not, where $m_j (1 \le j \le k)$ is the length of
 $a_j$'s.
 Finally, it checks whether
 $m_1 m_j = m_{j + 1} ~~(2 \le j \le k - 1)$ or not.
\end{sketchofproof}

\subsection{2Q1CA for $\boldsymbol{\{a^n b^{2^n} \mid n \ge 1\}}$}
\begin{lem}\label{power}
 There exists a 2R1CA $M_{b^{2^n}}$ which recognizes
 $\{b^{2^n} \mid n \ge 0\}$.
\end{lem}
\begin{proof}
 Let the state set $Q = \{q_0, q_1, q_2, q_3, q_4, q_5, q_6, q_7, q_8\}$,
 $\Qacc = \{q_7\}$ and $\Qrej = \{q_8\}$.
 For each $q \in Q$, $\sigma \in \Gamma$, $s \in \{0, 1\}$, we define
 the transition matrices $\{V_{\sigma, s}\}$, the counter function $C$,
 and the head function $D$ as follows:
\iffalse
 {\small
 \begin{eqnarray*}
  \begin{array}{lll}
   \begin{array}{l}
    \trans{\cent, \*}{0}{1},\\
    \trans{\cent, 1}{1}{2},\\
    \\
    C(q_0, a)     = +1,\\
    C(q_1, \cent) = +1,\\
    C(q_2, \cent) = -1,\\
    C(q_2, a)     = -1,\\
    C(q_5, a)     = +1,\\
    C(q, \sigma)  =  0 \quad \mbox{otherwise},\\
   \end{array} &
   \begin{array}{l}
    \trans{\dollar, 0}{3}{6},\\
    \trans{\dollar, \*}{2}{7},\\
    \\
    D(q_0) = \leftarrow,\\
    D(q_2) = \rightarrow,\\
    D(q_3) = \rightarrow,\\
    D(q_4) = \leftarrow,\\
    D(q)   = \downarrow \\
    \qquad \qquad \mbox{otherwise},\\
   \end{array} &
   \begin{array}{l}
    \trans{a, 0}{2}{3},\\
    \trans{a, 0}{3}{4},\\
    \trans{a, 0}{4}{0},\\
    \trans{a, 1}{2}{2},\\
    \trans{a, 1}{0}{5},\\
    \trans{a, 1}{5}{1},\\
    \\
    \\
    \\
   \end{array}
  \end{array}
 \end{eqnarray*}
 }
\fi
 {\small
 \begin{eqnarray*}
  \begin{array}{lll}
   \begin{array}{l}
    \trans{\cent,  0}{0}{0},\\
    \trans{\cent,  0}{3}{7},\\
    \trans{\cent, \*}{2}{8},\\
    \\
    C(q_1, \dollar) = +1,\\
    C(q_2, \dollar) = +1,\\
    C(q_2,       b) = -1,\\
    C(q_5,       b) = +1,\\
    C(q_6,       b) = +1,\\
    C(q  , \sigma ) =  0 \quad \mbox{otherwise},\\
   \end{array} &
   \begin{array}{l}
    \trans{\dollar, 0}{0}{1},\\
    \trans{\dollar, 1}{1}{2},\\
    \trans{\dollar, 1}{6}{1},\\
    \\
    D(q_0) = \rightarrow,\\
    D(q_2) = \leftarrow,\\
    D(q_3) = \leftarrow,\\
    D(q_4) = \rightarrow,\\
    D(q_6) = \rightarrow,\\
    D(q  ) = \downarrow \quad \mbox{otherwise},\\
   \end{array} &
   \begin{array}{l}
    \trans{b, 0}{0}{0},\\
    \trans{b, 0}{2}{3},\\
    \trans{b, 0}{3}{4},\\
    \trans{b, 0}{4}{6},\\
    \trans{b, 1}{2}{2},\\
    \trans{b, 1}{5}{6},\\
    \trans{b, 1}{6}{5},\\
    \\
    \\
    \\
   \end{array}
  \end{array}
 \end{eqnarray*}
 }

 Reversibility of this automaton can be checked easily.
\end{proof}

\begin{prop}
 Let $L_{\rm power}$ be $\{a^n b^{2^n} \mid n \ge 1\}$.
 For an arbitrary fixed positive integer $N \ge 2$,
 there exists a 2Q1CA $M_{\rm power}$ which
 accepts $x \in L_{\rm power}$ with probability $1$ and
 rejects $x \not\in L_{\rm power}$ with probability
 $1 - 1/N$. In either case, $M_{\rm power}$ halts after $O(N|x|)$ steps
 with certainty.
\end{prop}
\begin{proof}
 We can construct $M_{\rm power}$ from $M_{b^{2^n}}$ constructed
 in the proof of Lemma 16
 by using the technique of path branching. Its behavior is as follows.

 First, the automaton checks whether the input is of the form $a^+ b^+$.
 Second, the computation branches into $N$ paths, each with amplitude
 $1 / \sqrt{N}$. Along the $i$th path, the automaton checks whether
 the number of $b$'s is a power of 2,
 behaving in the same manner as $M_{b^{2^n}}$
 except that it remains stationary for $N - i + 1$ steps
 every time the tape head reads the symbol $\dollar$.
 Third, if the tape head reads the
 symbol $a$, then it remains for $i$ steps and after that moves to left.
 Finally, upon reading the symbol $\cent$, each computation path again
 splits according to the Discrete-Fourier Transformation, which yields the
 single accepting state and other rejecting states.

 By the construction of the automaton, we can easily see that,
 for $x \in L_{\rm power}$, the accepting state is entered with
 probability $1$, while for $x \not\in L_{\rm power}$,
 the accepting state is entered with probability
 $1 - 1/N$.
\end{proof}

\section{Conclusion and Open Problems}
In this paper, we gave the definition of 2Q1CAs and proved that
2Q1CAs is at least as powerful as 2D1CAs.
We also proved that 2Q1CAs recognize non-context-free languages,
such as
$\{a^n b^{n^2} \mid n \ge 1\}$,
$\{a^m b^n c^{mn} \mid m, n \ge 1\}$,
$\{a^n b^{2^n} \mid n \ge 1\}$, and so on.

Since these languages do not seem to be recognized by 2D1CAs,
our results may suggest that 2Q1CAs are more powerful than 2D1CAs.

As for two-counter cases,
it is known that two-way deterministic two-counter automata (2D2CAs)
can simulate deterministic Turing machines \cite{Minsky1961}.
From the result of Deutsch \cite{Deutsch1985},
quantum Turing machines cannot be beyond the undecidability.
Thus we can conclude that
two-way quantum two-counter automata (2Q2CAs)
are exactly as powerful as 2D2CAs in view of language recognition,
since we have a reversible simulation of a 2D2CA
by extending our technique of Theorem 12.
This suggests the possibility that,
even in the one-counter cases,
2Q1CAs are at most as powerful as 2D1CAs
in view of language recognition.
Thus it is actually open
whether 2Q1CAs are {\it strictly} more powerful than 2D1CAs.

As for simple models, we proved that simple 2D1CAs are as powerful as
2D1CAs. However, it remains open whether simple 2Q1CAs are as
powerful as 2Q1CAs or not.

Another interesting open problem is
to determine whether there are non-context-sensitive languages
that can be recognized by 2Q1CAs with bounded error.

\end{document}